\documentclass[twoside]{dis09}
\usepackage[latin1]{inputenc}
\usepackage[dvips]{graphicx,epsfig,color}
\usepackage{wrapfig,rotating}
\usepackage{amssymb,amsmath,array}

\begin{document}
\title{Electroweak Precision Data and New Gauge Bosons}

\author{Jens Erler
%
%
\vspace{.3cm}\\
%
Departamento de F\'isica Te\'orica, Instituto de F\'isica, \\
Universidad Nacional Aut\'onoma de M\'exico, 04510    M\'exico D.F., M\'exico
%
}

\maketitle

\begin{abstract}
I review constraints on the Standard Model (SM) Higgs boson from high energy electroweak (EW) precision data. The same data set also strongly limits various mixing effects of hypothetical extra neutral gauge bosons  ($Z'$) with the ordinary $Z$.  I also discuss low energy precision measurements which are sensitive to other aspects of $Z'$ physics, such as the direct exchange amplitude and the flavor or CP violating sectors. 
\end{abstract}

\section{High energy data: SM Higgs boson and $W$-$Z$-$Z'$ interdependence}
\subsection{SM global fit}

Of the four fundamental parameters of the EW sector of the SM --- the two gauge couplings, $g$ and $g'$ of $SU(2)_L$ and $U(1)_Y$, as well as the mass parameter and the self-coupling in the scalar Higgs potential--- three combinations are known to very high accuracy: (i) the Fermi constant, $G_F$, from the $\mu^\pm$ lifetime, $(\sqrt{2} G_F)^{-1/2} = 246.2209 \pm 0.0005$~GeV; (ii) the fine structure constant, $\alpha$, from the electron anomalous magnetic moment, $\alpha^{-1} = 137.035999679 (94)$; and (iii) the $Z$ boson mass, $M_Z = 91.1876 \pm 0.0021$~GeV, from the $Z$ line-shape scan at LEP~1~\cite{EWWG:2005ema}. The remaining parameter is the physical Higgs boson mass, $M_H$, which affects EW precision data only at the loop level.  Here the two most important observables are the $W$ boson mass, $M_W$, and the weak mixing angle, $\theta_W$. They provide two independent equations for the unknown $M_H$,
$$ \bar{\rho} \sin^2\bar\theta_W \cos^2\bar\theta_W (1 - \Delta \bar{r}) = 0.167145 (8) = \sin^2\theta_W \cos^2\theta_W (1 - \Delta r), $$
which enters the radiative correction parameters, $\bar\rho$, $\Delta r$, and $\Delta \bar{r}$. $\sin\bar\theta_W \equiv \bar{g}'/\sqrt{\bar{g}^2 + \bar{g}'^2}$ defines the weak mixing angle in the $\overline{\rm MS}$ renormalization scheme while $\cos\theta_W \equiv M_W/M_Z$ gives the on-shell definition. There is additional $M_H$-dependence in the total $Z$ width, $\Gamma_Z$, the $Zb\bar{b}$ vertex, and the low energy neutral current $\rho$ parameter.

\begin{wraptable}{l}{0.54\columnwidth}
\centerline{\begin{tabular}{|l|c|c|}
\hline
 & global fit & dominated by \\\hline
$m_t$ & $173.1 \pm 1.4$ & CDF and D\O \\\hline
$M_W$ & 80.380(15) & LEP 2, CDF and D\O \\\hline
$M_Z$ & 91.1874(21) & LEP 1 \\\hline
$\sin^2\bar\theta_W$& 0.23119(13) & $A_{FB}(b)$ and $A_{LR}$ \\\hline
$M_H$ & $96^{+29}_{-25}$ & $\sin^2\bar\theta_W$ and $M_W$ \\\hline
$\bar\alpha_s$ & 0.1185(16) & $Z$ and $\tau$ decays \\
\hline
\end{tabular}}
\caption{SM global fit results.}
\label{tab:SMfit}
\end{wraptable}

Table~\ref{tab:SMfit} summarizes the result of a global fit to all EW precision data. Shown are the masses of the top quark~\cite{TEVEWWG:2009ec} and the heavy bosons in GeV, as well as $\sin^2\bar\theta_W$ and the strong coupling constant (both in the $\overline{\rm MS}$ scheme evaluated at the $M_Z$ reference scale). The quoted uncertainties are from the SM fit parameters. The quality of the fit is rather good where $\chi^2_{\rm min} = 48.0$ is obtained for a number of effective degrees of freedom of 45 corresponding to a probability for a larger $\chi^2_{\rm min}$ of 35\%. The largest SM deviation is the $\sim 3\sigma$ effect in the muon anomalous magnetic moment~\cite{Bennett:2004pv}. In addition, it should be kept in mind that the forward-backward asymmetry into $b\bar{b}$ final states, $A_{FB}(b)$ (LEP~1), and the polarization asymmetry, $A_{LR}$ (SLD), deviate from each other at the $3\sigma$ level~\cite{EWWG:2005ema}, as well. 

\subsection{SM Higgs boson mass}

\begin{wrapfigure}{r}{0.64\columnwidth}
\centerline{\includegraphics[width=0.59\columnwidth]{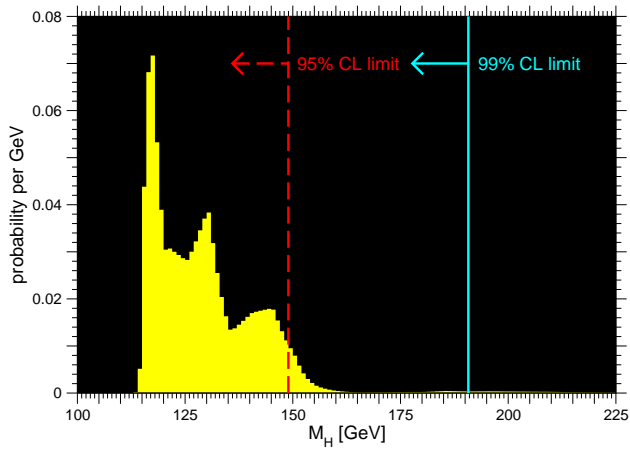}}
\caption{Probability distribution of $M_H$.}
\label{Fig:mh}
\end{wrapfigure}
The central 90\% CL range for $M_H$ derived indirectly from the EW precision data is,
$$ 58 \mbox{ GeV} < M_H < 146  \mbox{ GeV}. $$
A significant fraction of this range is excluded by direct searches at LEP~2~\cite{Barate:2003sz}, where the  bound, $M_H > 114.4 \mbox{ GeV}$, has been obtained at the 95\% CL. One can combine the direct and indirect results, but to do so one has to add the full LEP~2 likelihood curve as a function of $M_H$ to the global fit $\chi^2$ function, yielding,
$$ 115 \mbox{ GeV} < M_H < 168  \mbox{ GeV}. $$
If one instead simply multiplies (as is sometimes done) by a Heaviside step function with the step located at the lower bound, one obtains combined results which are biased towards higher $M_H$ values. Finally, one can add the search results from the Tevatron~\cite{TEVEWWG:2009pt}, as well, which by themselves eliminate the 95\% CL range, $160 \mbox{ GeV} < M_H < 170  \mbox{ GeV}$, and which gives the 90\% CL range allowed by {\em all} data,
$$ 115 \mbox{ GeV} < M_H < 149  \mbox{ GeV}. $$
Figure~\ref{Fig:mh} shows the probability density for $M_H$ assuming the SM and including all data~\cite{url}.

\subsection{Universal $Z'$ bosons}

Among the most widely studied $Z'$ bosons are the $Z_\chi$ and the $Z_\psi$, defined as the gauge bosons associated with $U(1)$ subgroups of Grand Unified Theories, namely $SO(10) \to SU(5) \times U(1)_\chi$ and  $E_6 \to SO(10) \times U(1)_\psi$. The combination, $Z_\eta = \sqrt{3/8}\ Z_\chi - \sqrt{5/8}\ Z_\psi$, appears in a class of heterotic string string compactifications on Calabi-Yau manifolds.  Similarly, the $Z_{LR}$ of left-right symmetric models is defined through, $SU(2)_R \times U(1)_{B-L} \to U(1)_Y \times U(1)'$, where $B-L$ refers to lepton minus baryon number. One can also define a $Z_R$ boson as the third component of $SU(2)_R$.  This possibility does not occur in popular models, but fits the data well. The sequential $Z_{\rm seq}$ is defined to couple like the ordinary $Z$ to the known fermions.  It is not expected to arise in field theories (unless it couples differently to exotic fermions) but could play the r$\hat{\rm o}$le of an excited $Z$ in models of compositeness or extra dimensions. As a final example, the $Z_X$ boson appears in a model motivated by seesaw neutrino masses~\cite{Adhikari:2008uc}.

Now one can investigate how $Z'$ bosons impact the high energy data. We are mostly interested in the $Z'$ boson itself and not in other aspects of some specific model containing it. If in this case its mass, $M_{Z'}$, approaches infinity one recovers the SM, a feature which differs from some other types of physics beyond the SM, such as a possible fourth fermion generation. There is a general relation~\cite{Langacker:1984dp} which follows from diagonalizing the neutral gauge boson mass matrix,
$$\tan^2\theta_{ZZ'} = {M_0^2 - M_Z^2\over M_{Z'}^2 - M_0^2},$$
where $M_0 = M_W/cos\theta_W$.  Additional modifications of the $W$-$Z$-$Z'$ interdependence can arise through higher dimensional Higgs representations at the tree level or through exotic fermions (usually necessary to cancel gauge anomalies) or scalar superpartners (in supersymmetric models) in loops. Both of these effects can be absorbed in the so-called oblique $T$ parameter~\cite{Peskin:1990zt} which one may allow. If the Higgs sector is known in a specific model, then there exists an additional constraint~\cite{Langacker:1991pg},
$$ \theta_{ZZ'} = C\ {g_2\over g_1} {M_Z^2\over M_{Z^\prime}^2}, $$
where $g_1 = g/\cos\theta_W$, $g_2$ is the $U(1)^\prime$ gauge coupling, and $C$ is a function of vacuum expectation values of the Higgs fields. If $\theta_{ZZ'} = 0$ then the high energy data are virtually blind to the $Z'$. Conversely, the high energy data usually constrain $\theta_{ZZ'}$ to the $10^{-3}$ level.

\section{Low energy data: $Z'$ mediated amplitude and non-universal $Z'$}
\subsection{Low energy constraints}

The low energy sector is much more sensitive to the specific $Z'$ couplings than the high energy data, and some of the strongest constraints arise from $\gamma Z'$ interference at very low momentum transfer, $Q^2$. A prime example is the nuclear spin independent part of the parity violating interaction measured in atomic physics. For a given nucleus, $X$,  it can be written as a sum of products of quark vector and electron axial-vector couplings, the so-called weak charge, $Q_W(X)$. The most precisely known one is $Q_W(^{133}{\rm Cs}) = -73.16 \pm 0.29 \mbox{ (experiment~\cite{Wood:1997zq})} \pm 0.20 \mbox{ (atomic theory~\cite{Porsev:2009pr})}$.  Unlike previously, this is in perfect agreement with the SM prediction, $Q_W(^{133}{\rm Cs}) = 188\ Q_W(u) + 211\ Q_W(d) = - 73.16$, written in terms of the up and down quark weak charges. These are shifted due to a $Z'$ by $\Delta Q_W(q) \propto (e_L - e_R) (q_L + q_R) M_Z^2/M_{Z'}^2$, where $e_{L,R}$ and $q_{L,R}$ are the left and right-handed $Z'$ couplings to charged leptons and quarks. $Q_W(^{133}{\rm Cs})$  by itself gives a 95\% CL limit, $M_\chi \geq 0.89$~TeV. By contrast, there is no constraint for the $Z_\psi$ since $SO(10)$ symmetry implies that all left-handed SM fields and anti-fields have the same charges, so that $q_L + q_R = 0$. 

The weak charge of the electron, $Q_W(e) =  \rho (-1 + 4 \kappa \sin^2\theta_W[\sqrt{Q^2}]) =-0.0472$ (SM), is defined analogously ($\rho$ and $\kappa$ collect the radiative corrections), and has been measured in polarized M\o ller scattering~\cite{Anthony:2005pm} at SLAC by directing the SLC $e^-$ beam with energies of 45 and 48~GeV and a polarization of  $89 \pm 4\%$ on a fixed hydrogen target.  The resulting low $Q^2 \approx 0.026$~GeV$^2$ implies a tiny right-left polarization asymmetry, $A_{RL} = - (1.31 \pm 0.14 \pm 0.10) \times10^{-7}$, and $Q_W(e) = - 0.0403 \pm 0.0053$ could be extracted. The shift due to a $Z'$ is given by $\Delta Q_W(e) \propto (e_L - e_R) (q_L + q_R) M_Z^2/M_{Z'}^2$ (for $\sin^2\theta_{ZZ'} = 0$), providing a 95\% CL limit, $M_\chi \geq 0.67$~TeV, while again there is no constraint for the $Z_\psi$ ($e_L + e_R = 0$). This constraint is manifestly complementary to LEP~2 (which was sensitive only to parity conserving observables) and the Tevatron (which lacks purely leptonic quantities altogether).

As for the future, the Qweak experiment at JLab is scheduled to measure the proton's weak charge,  $Q_W(p)$, in elastic $ep$ scattering at the 6 GeV CEBAF beam to $\pm 0.0029$ precision.  Parity violating deep inelastic electron scattering with both the 6 and 12 GeV CEBAF beam may determine the combination $2 C_{1u} - C_{1d} + 0.84 (2 C_{2u} - C_{2d})$ to $\pm 0.0049$ precision, where the $C_{2q}$ are defined like the $C_{1q}$ with vector and axial-vector couplings interchanged. This is assuming the absence of higher twist and charge symmetry violation effects in the parton distribution functions, which can be studied by varying the kinematic variables, $Q^2$ and $x$, respectively. Finally, the upgraded CEBAF beam to 12~GeV can be used to remeasure $A_{RL}$ with five times greater precision which would yield a determination of the weak mixing angle of $\pm 0.00029$, an accuracy similar to the worlds most precise measurements. After a possible discovery this measurement would also be important to break a degeneracy in the $Z'$ coupling space~\cite{Petriello:2008zr}. The expected 95\% CL limits of these three experiments are $M_\chi \geq 0.67$~TeV, 0.45~TeV and 1.07~TeV, respectively, with no $Z_\psi$ constraints.

Some observables are affected only at the loop level. These effects are usually very small, but not necessarily negligible, and in some cases they are finite and can be straightforwardly taken into account. An example is the unitarity condition on
the first row of the CKM matrix. Using the values, $|V_{ud}| = 0.97425 \pm 0.00008 \mbox{ (experiment)} \pm 0.00010 \mbox{ (nuclear structure)} \pm 0.00008 \mbox{ (radiative corrections~\cite{Marciano:2005ec})} = 0.97425 \pm 0.00022$ from super-allowed $0^+ \to 0^+$ nuclear $\beta$ decays~\cite{Hardy:2008gy}, $|V_{us}| = 0.22478 \pm 0.00124$ from $K_{\ell 3}$ decays~\cite{Bossi:2008aa}, and $|V_{us}/V_{ud}| = 0.23216 \pm 0.00145$ from $K_{\ell 2}$ decays~\cite{Bossi:2008aa}, one obtains, $|V_{ud}|^2 + |V_{us}|^2 + |V_{ub}|^2 = 1.0000 \pm 0.0006$, in perfect agreement (unlike previously) with the SM and unitarity. The presence of a $WZ'$ box diagram would upset the unitarity relation (when normalized to $G_F$) by an amount $\propto e_L (e_L - q_L) \ln (M_{Z'}/M_W)/(M_{Z'}^2/M_W^2 - 1)$~\cite{Marciano:1987ja}, from which we obtain the 95\% CL limit, $M_\chi \geq 265$~GeV (again there is no $Z_\psi$ constraint because $e_L = q_L$). 

\subsection{$Z'$ bosons: results}

\begin{wrapfigure}{r}{0.49\columnwidth}
\centerline{\includegraphics[width=0.44\columnwidth]{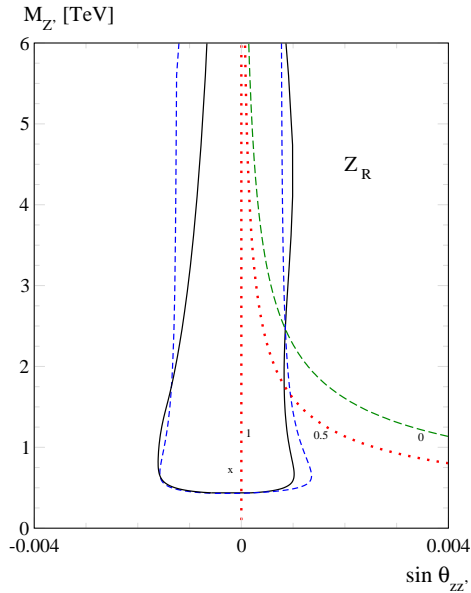}}
\caption{Exclusion curves for the $Z_R$.}
\label{Fig:ZR}
\end{wrapfigure}

Figure~\ref{Fig:ZR} shows 90\% CL exclusion contours for the $Z_R$ case. The solid (black) line uses the constraint $\rho_0 = 1$ while the dashed (blue) line is for $\rho_0$ free. We also show the extra constraints for some specific Higgs sectors. These are indicated by the dotted (red) lines except for a scenario preferred by supersymmetry shown as a long-dashed (green) line. The best fit location (for $\rho_0 = 1$) is indicated by an {\tt "x"}. The $Z_R$ actually gives a rather good fit and technically even a 90\% CL {\em upper limit\/} $M_{Z'} < 29$~TeV can be set in this case.
\begin{wraptable}{l}{0.39\columnwidth}
\centerline{\begin{tabular}{|l|r|r|r|}
\hline
                  & EW & CDF & LEP~2 \\\hline
$Z_\chi$  &   1,141 &   892 &     781~\cite{Abbiendi:2003dh} \\\hline
$Z_\psi$  &      147 &   878 &     481~\cite{Alcaraz:2006mx}  \\\hline
$Z_\eta$  &      427 &    982 &    515~\cite{Abbiendi:2003dh} \\\hline
$Z_{LR}$ &      998 &    630 &    804~\cite{Alcaraz:2006mx} \\\hline
$Z_{\rm seq}$ & 1,403 & 1,030 & 1,787~\cite{Alcaraz:2006mx} \\
\hline
\end{tabular}}
\caption{Lower mass limits for selected $Z'$ bosons in GeV.}
\label{tab:Zprime}
\end{wraptable}

Table~\ref{tab:Zprime} shows the 95\% CL lower mass limits for various $Z'$ bosons from the EW precision data (for $T = 0$), compared to collider results from CDF~\cite{Aaltonen:2008ah} (assuming no exotic or supersymmetric decay channels are kinematically allowed) and LEP~2~\cite{Alcaraz:2006mx} (in general for $\sin\theta_{ZZ'} = 0$, but the OPAL Collaboration~\cite{Abbiendi:2003dh} allowed and constrained $\sin\theta_{ZZ'} = 0$, as well). We emphasize that these limits are highly complementary (for more details, see Ref.~\cite{Erler:2009jh}).

Conceivably, the $Z'$ boson couples in a family non-universal way. In that case, usually unacceptably large effects of charged lepton flavor violation, CP violation and new flavor changing neutral currents may be induced through (i) intergenerational couplings (charges) of the $Z'$, (ii) fermion mass matrices even in the case of a diagonal $Z'$,  (iii) the ordinary $Z$ if $\sin\theta_{ZZ'} \neq 0$, and (iv) the mixing of ordinary fermions and exotics (through both, the $Z$ and the $Z'$). The effects are difficult to quantify even within a given model. But it has been shown~\cite{Langacker:2000ju} that the contributions to neutral $B$ and $B_s$ mixing and to CP violation in the $K$ system are typically too large even for equal $U(1)'$ charges of the first two families. For an EW fit in the presence of a non-universal $Z'$, see Ref.~\cite{Erler:1999nx}. 

\section*{Acknowledgments}
It is a pleasure to thank Paul Langacker, Shoaib Munir and Eduardo Rojas for collaboration. 
This work is supported by CONACyT (M\'exico) project 82291--F. 
 

\begin{footnotesize}

\end{footnotesize}

\begin{thebibliography}{99}

\bibitem{url} Slides: \\
\verb$http://indico.cern.ch/contributionDisplay.py?contribId=35&sessionId=2&confId=53294$

\bibitem{EWWG:2005ema}
ALEPH, DELPHI, L3, OPAL and SLD Collaborations, LEP EW Working Group and SLD EW and Heavy Flavour Groups:
S. Schael {\it et al.}, Phys.\ Rept.\  {\bf 427}, 257 (2006).

\bibitem{TEVEWWG:2009ec}
Tevatron EW Working Group and CDF and D\O\ Collaborations: arXiv:0903.2503 [hep-ex].

\bibitem{Bennett:2004pv}
Muon g-2 Collaboration: G.W.~Bennett {\it et al.}, Phys.\ Rev.\ Lett.\  {\bf 92}, 161802 (2004).

\bibitem{Barate:2003sz}
LEP Working Group for Higgs boson searches and ALEPH, DELPHI, L3 and OPAL Collaborations: R.~Barate {\it et al.},                           Phys.\ Lett.\  {\bf B565}, 61 (2003).

\bibitem{TEVEWWG:2009pt}
Tevatron New Phenomena and Higgs Working Group, CDF and D\O: arXiv:0903.4001 [hep-ex].

\bibitem{Adhikari:2008uc}
R.~Adhikari, J.~Erler and E.~Ma, Phys.\ Lett.\  {\bf B672}, 136 (2009).

\bibitem{Langacker:1984dp}
P.~Langacker, Phys.\ Rev.\  {\bf D30}, 2008 (1984).

\bibitem{Peskin:1990zt}
M.E.~Peskin and T.~Takeuchi, Phys.\ Rev.\ Lett.\  {\bf 65}, 964 (1990).

\bibitem{Langacker:1991pg}
P.~Langacker and M.~Luo, Phys.\ Rev.\  {\bf D45}, 278 (1992).

\bibitem{Wood:1997zq}
C.S.~Wood, {\it et al.}, Science {\bf 275}, 1759 (1997).

\bibitem{Porsev:2009pr}
S.G.~Porsev, K.~Beloy and A.~Derevianko, Phys.\ Rev.\ Lett.\  {\bf 102}, 181601 (2009).

\bibitem{Anthony:2005pm}
SLAC E158 Collaboration: P.L.~Anthony {\it et al.}, Phys.\ Rev.\ Lett.\  {\bf 95}, 081601 (2005).

\bibitem{Petriello:2008zr}
F.~Petriello and S.~Quackenbush, Phys.\ Rev.\  {\bf D77}, 115004 (2008).

\bibitem{Marciano:2005ec}
W.J.~Marciano and A.~Sirlin, Phys.\ Rev.\ Lett.\  {\bf 96}, 032002 (2006).

\bibitem{Hardy:2008gy}
J.C.~Hardy and I.S.~Towner, Phys.\ Rev.\  {\bf C79}, 055502 (2009).

\bibitem{Bossi:2008aa}
KLOE Collaboration: F.~Bossi {\it et al.}, Riv.\ Nuovo Cim.\  {\bf 031}, 531 (2008).

\bibitem{Marciano:1987ja}
W.J.~Marciano and A.~Sirlin, Phys.\ Rev.\  {\bf D35}, 1672 (1987).

\bibitem{Aaltonen:2008ah}
CDF Collaboration: T.~Aaltonen {\it et al.}, Phys.\ Rev.\ Lett.\  {\bf 102}, 091805 (2009).

\bibitem{Alcaraz:2006mx}
ALEPH, DELPHI, L3 and OPAL Collaborations and LEP EW Working Group: arXiv:hep-ex/0612034.

\bibitem{Abbiendi:2003dh}
OPAL Collaboration: G.~Abbiendi {\it et al.}, Eur.\ Phys.\ J.\  {\bf C33}, 173 (2004).

\bibitem{Erler:2009jh}
J.~Erler, P.~Langacker, S.~Munir and E.~Rojas, arXiv:0906.2435 [hep-ph], to appear in JHEP.

\bibitem{Langacker:2000ju}
P.~Langacker and M.~Pl\"umacher, Phys.\ Rev.\  {\bf D62}, 013006 (2000).

\bibitem{Erler:1999nx}
J.~Erler and P.~Langacker, Phys.\ Rev.\ Lett.\  {\bf 84}, 212 (2000).

\end{thebibliography}
\end{document}